\documentclass[preprint2,times]{aastex631}

\newcommand{\kms}{km $\mathrm{s^{-1}}$ }

\usepackage{color}
\usepackage{multirow}
\usepackage{threeparttable}
\usepackage{ulem}
\usepackage{pifont}

\begin{document}

\title{Absence of High-mass Prestellar Cores in the Orion Giant Molecular Cloud}

\author{Wenyu Jiao}
\affiliation{Kavli Institute for Astronomy and Astrophysics, Peking University, Haidian District, Beijing 100871, People’s Republic of China}
\affiliation{Department of Astronomy, School of Physics, Peking University, Beijing, 100871, People's Republic of China}
\affiliation{Shanghai Astronomical Observatory 
Chinese Academy of Sciences, 80 Nandan Road, Shanghai 200030, People’s Republic of China}
\author{Ke Wang}
\affiliation{Kavli Institute for Astronomy and Astrophysics, Peking University, Haidian District, Beijing 100871, People’s Republic of China}
\author{Fengwei Xu}
\affiliation{Kavli Institute for Astronomy and Astrophysics, Peking University, Haidian District, Beijing 100871, People’s Republic of China}
\affiliation{Department of Astronomy, School of Physics, Peking University, Beijing, 100871, People's Republic of China}
\affiliation{I. Physikalisches Institut, Universität zu Köln, Zülpicher Str. 77, D-50937 Köln, Germany}

\correspondingauthor{Ke Wang}
\email{kwang.astro@pku.edu.cn}


\begin{abstract}

A fundamental difference between ``core-fed'' and ``clump-fed'' star formation theories lies in the existence or absence of high-mass cores at the prestellar stage.
However, only a handful of such cores have been observed. Here, different than previous search in distributed star formation regions in the Galactic plane, we search for high-mass prestellar cores in the Orion GMC, by observing the 7 most massive starless cores selected from previous deep continuum surveys.
We present ALMA ACA Band 6 and Band 7 continuum and line observations toward the 7 cores, in which we identify 9 dense cores at both bands. The derived maximum core mass is less than $11\, M_{\odot}$, based on different dust temperatures. We find no high-mass prestellar cores in this sample, aligning with the results of previous surveys, thereby challenging the existence of such cores in Orion. Outside Orion, further detailed studies are needed for remaining high-mass prestellar core candidates to confirm their status as massive, starless cores.

\end{abstract}

\keywords{Dust Continuum Emission (412); Massive stars (732); Molecular clouds(1072); Star formation (1569)}

\section{Introduction} \label{sec:intro}

\par High-mass stars are rare \citep{Kroupa_2002}, but they dominate the energy budget of galaxies and drive galaxy evolution through strong stellar feedback and element enhancement. Therefore, massive star formation is fundamentally important not only for itself, but also for determining star formation rate and initial mass function for extragalactic studies. However, unlike low-mass stars, the mechanism behind the formation of high-mass stars remains under debate \citep{Tan_2014,Motte_2018}.

One of the breaking points on the observational side is to pinpoint the initial condition of massive star formation by taking efforts on the high resolution observations \citep{Zhang_2009,qz11,qz15,me11,Beuther_2013,Wang_2014,Feng2016,Svoboda_2019,Sanhueza_2019,Morii2023}. Following the path of theoretical or simulation works, two types of initial conditions are proposed. One is the ``competitive accretion'' (CA) model \citep{Bonnell_2001}, which is a dynamical, ``quick'' process where Jeans-like cores (typical $\mathrm{M_{Jeans}} \sim 1 \ \mathrm{M_{\odot}}$) continue to grow via gas accretion while the embedded stellar embryos grow. This scenario is supported by some observations, where fragmentation at smaller scales and embedded star-forming activities, such as outflows, are identified (e.g., \citealt{Tan_2016,Feng2016,Cyganowski_2022,Jiao_2023}). More importantly, studies across evolutionary stages have found the core mass growth and dense gas assembly along the evolution of massive star-forming regions, through clump-fed gas infall \citep{Baobab2012_G10.6, YuanJH2018_G22, ZhouJW2022,ZhangSJ23-ClumpFed,Xu_2024,Xu2024quarksII}, filament mediated gas flows \citep{qz15,WangKe2018}, 
and accretion evident by outflows \citep{me11,WangKe2012,Wang_2014,Baug2021,Jiao_2023}. That means the massive cores do not need to be massive enough at the initial stage, but can obtain its gas subsequently. 

In contrast, the ``turbulent core accretion'' (TA) model \citep{Mckee_Tan_2003} scenario is related to a “slow” monolithic collapse of pre-assembled massive cores (typically 100 $\mathrm{M_\odot}$). Till now, only few high-mass prestellar core candidates (HMPCs) are known: G11P6-SMA1 in the ``Snake'' infrared dark cloud \citep{Wang_2014}, C2c1a in the ``Dragon'' infrared dark cloud \citep{Barnes_2023}, W43-MM1$\#$134 \citep{Nony_2023}, W43-MM1$\#$6 \citep{Nony_2018} in the W43 starburst region, and CygXN53-MM2 in the relatively nearby Cygnus-X complex \citep{Bontemps_2010}. 

\begin{figure*}[bt!]
    \centering
    \includegraphics[width=1.0\linewidth]{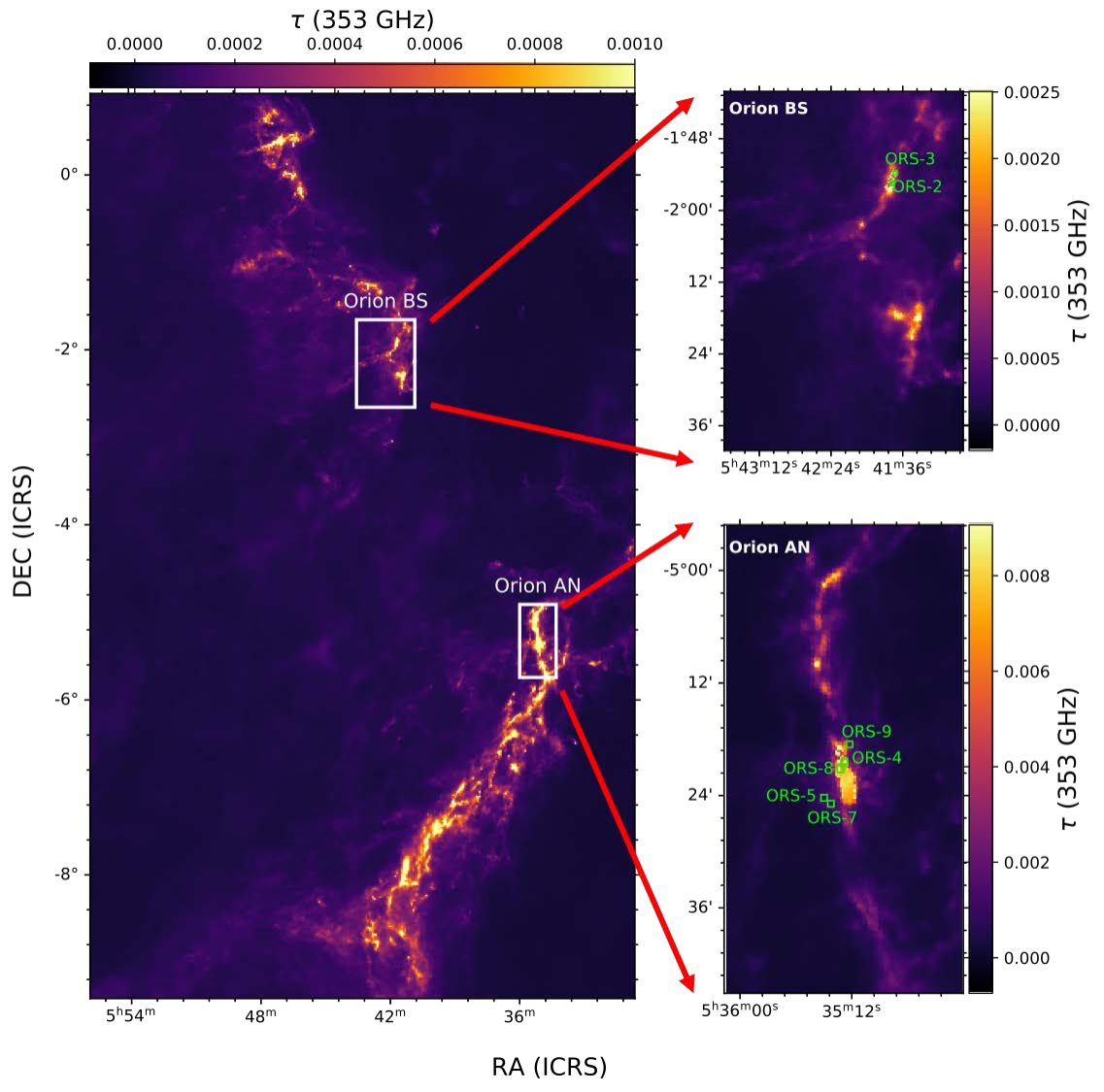}
    \caption{Spatial distribution of seven cores in Orion. The background is the 353 GHz opacity retrieved from \citet{Lombardi_2014}. Two small regions that covers our sample are further zoomed in. The green box represents the position of each core.}
    \label{fig:cores}
\end{figure*}



\subsection*{O\lowercase{ur} S\lowercase{trategy}}

In this work, we set out to carry out a systematic search for high-mass prestellar cores in a single giant molecular cloud (GMC), the Orion GMC.
This strategy is more efficient than previous searches in targeted regions (Section \ref{sec:intro}), and systematic Galaxy-wide searches \citep[e.g.,][]{YuanJH17-HMSC,HuangB23-HMSC,WangKe_sbmt}.

Orion is the nearest \citep[d= 414 pc]{Menten_2007} and the most well-studied high-mass star-forming region.
Dense cores at early evolutionary stages in the Orion nebula have been investigated by both single-dish telescope surveys (e.g., \citealt{Salji_2015,Konyves_2020}) and interferometric observations (e.g., \citealt{Dutta_2020,Sahu_2021}). 
Our sample selection starts from the work of \cite{Nutter_2007}, who made a census of 393 dense cores in the Orion star-forming regions using archival deep JCMT/SCUBA 450/850 $\mu$m imaging. To exclude young stellar objects, starless cores were classified by comparing them with $Spitzer$ images. We further excluded young protostars identified in the Herschel Orion Protostar Survey \citep{Furlan_2016}. Finally, we selected cores with mass larger than $\sim$30 $\mathrm{M_{\odot}}$ at a constant dust temperature of 20 K (except one core with 24.4 $\mathrm{M_{\odot}}$). As a result, our sample is comprised of seven cores named ORion Sources (hereafter ORS). The spatial distribution of these seven cores is shown in Figure \ref{fig:cores}. Two of them are located in Orion B and the rest lies in Orion A. Their mass-size diagram is plotted in Figure \ref{fig:sample} as compared to other known HMPCs. We then use ALMA to resolve these candidate HMPCs to investigate their internal structure.

\section{Observations} \label{sec:obs}

The seven sources were observed with the 7-m Atacama Compact Array (ACA; also known as the Morita Arrary, \citealt{Iguchi_2009}) in Cycle 7 (Project ID: 2019.2.00094.S, PI: Ke Wang). The observations were performed in Band 6 on December 23, 2019 and Band 7 on January 8, 2020. J0538-4405 was used as the bandpass and flux calibrator. The phase calibrator was J0607-0834. For the Band 6 observations, the maximum recoverable scale was $28.6^{\prime \prime}$ and the primary beam size was $44.6^{\prime \prime}$. The integration time was approximately 5 minutes for each source. The observations included two wide spectral windows with a bandwidth of 2 GHz centered at 216.68 and 230.89 GHz for continuum and lines, including $\mathrm{DCO^+}$ ($J$=$3-2$), SiO ($J$=$5-4$), DCN ($J$=$3-2$), and CO ($J$=$2-1$). Additionally, there were five narrow spectral windows with a velocity resolution of $\sim$ 0.16 \kms for molecular lines, including $\mathrm{H_2CO}$ ($J_{\mathrm{K}{\mathrm{a}},\mathrm{K}{\mathrm{c}}}$=$3_{0,3}-2_{0,2}$), $\mathrm{CH_3OH}$ ($J_{K}=4_{2}-3_{1}$), $\mathrm{H_2CO}$ ($J_{\mathrm{K}{\mathrm{a}},\mathrm{K}{\mathrm{c}}}$=$3_{2,2}-2_{2,1}$), $\mathrm{C^{18}O}$ ($J$=$2-1$), $^{13}\mathrm{CS}$ ($J$=$5-4$), and $\mathrm{N_2D^{+}}$ ($J$=$3-2$). For the Band 7 observations, the maximum recoverable scale was $28.6^{\prime \prime}$ and the primary beam size was $35.2^{\prime \prime}$. The integration time was approximately 10 minutes for each source. The observations had two wide spectral windows with a bandwidth of 2 GHz centered on 278.84, 288.77 GHz for continuum and three narrow spectral windows with a velocity resolution of $\sim$ 0.13 \kms for molecular lines including $\mathrm{N_2H^+}$ ($J$=$3-2$), $\mathrm{DCO^+}$ ($J$=$4-3$), $\mathrm{DCN}$ ($J$=$4-3$). We summarize our spectral window setting in Table \ref{line information}. 
\par Data reduction was performed using CASA software package version 5.6.1 \citep{McMullin_2007}. For continuum data, we manually selected line-free channels, and the continuum image was obtained using the $tclean$ task by averaging line-free channels with a Briggs’s robust weighting of 0.5 to the visibilities. The average synthesized beam sizes are $8.3^{\prime \prime} \times 4.2^{\prime \prime}$ ($\sim 2400$ AU at 414 pc distance) with a position angle of $-88^{\circ}$ for Band 6 and $6.5^{\prime \prime} \times 3.3^{\prime \prime}$ ($\sim 1900$ AU at 414 pc distance) with a position angle of $87^{\circ}$ for Band 7, respectively. The image sizes are $81^{\prime \prime} \times 81^{\prime \prime}$ and $62^{\prime \prime} \times 62^{\prime \prime}$ for Band 6 and Band 7, respectively. The RMS noise of the continuum image is measured in an emission-free region, and the detailed values of every source are shown in Table \ref{observed parameter}. For spectral line, the line cube was cleaned using the $tclean$ task after removing the continuum emission using the $uvcontsub$ task, with a Briggs robust weighting of 0.5. We use the auto-multithresh algorithm with parameters set to the standard values for 7 m data, as specified in the user guides\footnote{see details in \url{https://casaguides.nrao.edu/index.php?title=Automasking Guide}}.

\begin{table*}
	\centering
	\caption{\centering{Summary of Spectral Line Information}}
	\label{line information}
	\begin{tabular}{m{1.5cm}m{3cm}m{2cm}m{1.5cm}m{3.5cm}m{2cm}m{2.2cm}ccccccc} 
		\hline
		\hline
Line &  \ \ \ \ \ Transition &  Rest Freq. & $E_u/k$ & Velocity Resolution & Beam Size$^{a}$ & RMS Level \\
 &  & $\ \ $ (GHz) & $ \ $(K)&$\ \ \ \ \ \ \  (\mathrm{km \ s^{-1}}$) & \ \ \  ($^{\prime\prime}\times ^{\prime\prime}$) & (Jy beam$^{-1}$) \\
\hline
Band 6& & && & & & \\
\hline
$\mathrm{DCO^{+}}$ & \ \ \ \ \ $J$=$3-2$	&  216.112580 &	 20.74 &  \ \ \ \ \ \ \ \ \ \ 2.71 &    \ \ 8.6$\times$4.6  & \ \ \ \ \ 0.03 \\
$\mathrm{SiO}$ & \ \ \ \ \ $J$=$5-4$	&  217.104980 &	 31.26 & \ \ \ \ \ \ \ \ \ \ 2.70 &   \ \ 8.6$\times$4.6 & \ \ \ \ \ 0.03\\
$\mathrm{DCN}$ & \ \ \ \ \ $J$=$3-2$	&  217.238530 &	 20.85  &\ \ \ \ \ \ \ \ \ \ 2.70 &   \ \ 8.6$\times$4.6  & \ \ \ \ \ 0.03\\
$\mathrm{H_2CO}$ & $J_{\mathrm{K}_{\mathrm{a}}, \mathrm{K}_{\mathrm{c}}}$=$3_{0,3}-2_{0,2}$	&  218.222192 &	20.96 &  \ \ \ \ \ \ \ \ \ \ 0.17 &  \ \ 8.6$\times$4.6  & \ \ \ \ \ 0.10\\
$\mathrm{CH_3OH}$ & \ \ $J_{K}=4_{2}-3_{1}$	&  218.440063 &	45.46 &  \ \ \ \ \ \ \ \ \ \ 0.17 &  \ \ 8.6$\times$4.6  & \ \ \ \ \ 0.11\\
$\mathrm{H_2CO}$ & $J_{\mathrm{K}_{\mathrm{a}}, \mathrm{K}_{\mathrm{c}}}$=$3_{2,2}-2_{2,1}$	&  218.475632 &	68.09 &  \ \ \ \ \ \ \ \ \ \ 0.17 &    \ \ 8.6$\times$4.6  & \ \ \ \ \ 0.10\\
$\mathrm{C^{18}O}$ & \ \ \ \ \ $J$=$2-1$	&  219.560358 &	 15.81  &\ \ \ \ \ \ \ \ \ \ 0.17 &  \ \ 8.6$\times$4.6  & \ \ \ \ \ 0.13 \\
$\mathrm{CO}$ & \ \ \ \ \ $J$=$2-1$	&  230.538000 &	 16.60  &\ \ \ \ \ \ \ \ \ \ 2.54 &  \ \ 8.1$\times$4.4 & \ \ \ \ \ 0.03\\
${{ }^{13} \mathrm{CS}}$ & \ \ \ \ \ $J$=$5-4$	&  231.220686 &	 33.29  &\ \ \ \ \ \ \ \ \ \ 0.17 &  \ \ 8.1$\times$4.4  & \ \ \ \ \ 0.13 \\
$\mathrm{N_2D^{+}}$ & \ \ \ \ \ $J$=$3-2$	&  231.321828 &	 22.20 &  \ \ \ \ \ \ \ \ \ \ 0.17 &  \ \ 8.1$\times$4.4 & \ \ \ \ \ 0.14 \\
\hline
Band 7 & & && & & & \\
\hline
$\mathrm{N_2H^{+}}$ & \ \ \ \ \ $J$=$3-2$	&  279.511749 &	 26.83 &  \ \ \ \ \ \ \ \ \ \ 0.13 &  \ \ 6.5$\times$3.5 & \ \ \ \ \ 0.10 \\
$\mathrm{DCO^{+}}$ & \ \ \ \ \ $J$=$4-3$	&  288.143858 &	 34.57 &  \ \ \ \ \ \ \ \ \ \ 0.13 &  \ \ 6.3$\times$3.4 & \ \ \ \ \ 0.10 \\
$\mathrm{DCN}$ & \ \ \ \ \ $J$=$4-3$	&  289.644917 &	 34.75 &  \ \ \ \ \ \ \ \ \ \ 0.13 &  \ \ 6.3$\times$3.4 & \ \ \ \ \ 0.10 \\
\hline
	\end{tabular}
    \begin{tablenotes}
        \footnotesize
        \item $^a$The synthesized beam size is the average value of each source.

      \end{tablenotes}
\end{table*}

\section{Results}
\label{sec:results}
\subsection{Core Identification}
\par
\begin{figure*}[bt!]
    \centering
    \includegraphics[width=1.0\linewidth]{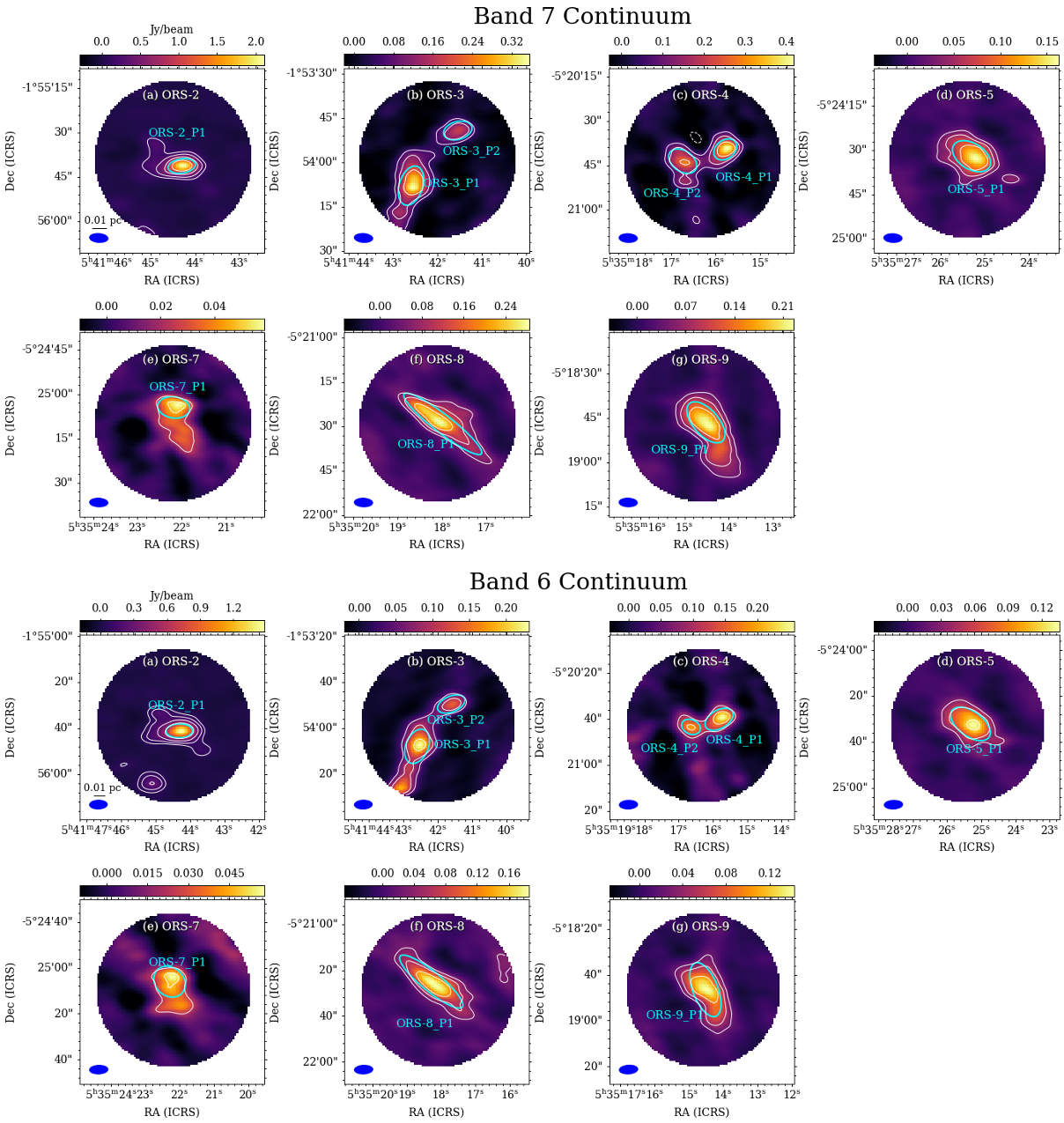}
    \caption{ACA 1.06 mm and 1.34 mm continuum images of seven sources shown in color before primary beam correction. The cyan ellipse represents the size of the identified cores convolved with beam. The white contours show continuum emission at levels of [-6, 6, 12, 24, 48]$\sigma$, where $\sigma$ can be found in Table \ref{observed parameter}. The blue ellipse in the bottom left marks the synthesized beam.}
    \label{fig:continuum}
\end{figure*}

 Figure \ref{fig:continuum} presents the continuum images without primary beam correction at 1.06 mm (Band 7; centered at 282.9 GHz) and 1.34 mm (Band 6; centered at 223.8 GHz). To extract the dense cores within the fields, we adopt a procedure that has been performed in previous studies (e.g., \citealt{qz15, Li_2019,LiuHL21,Xu2023}). We firstly use the $Dendrogram$ algorithm\footnote{\url{https://dendrograms.readthedocs.io/en/stable/}} on the continuum image without primary beam correction to identify dense cores. Because we focus on the cores with high mass, the \emph{min\_value} is set to be 6$\sigma$, where $\sigma$ is the RMS noise of the continuum image. The \emph{min\_delta} is set to be 3$\sigma$ and the \emph{min\_npix} equals to the number of pixels within the beam area. To accurately derive the physical parameters of these cores, we make use of the $imfit$ function in CASA on the continuum image after primary beam correction. Because of the different image sizes and sensitivities at two different bands, we only select the cores identified at both two bands. We identify nine dense cores in the seven sources, while three cores are excluded because they are only identified in single band. We find multiple systems in ORS-3 and ORS-4, and a filamentary structure in ORS-8. The detailed physical properties of identified cores are shown in Table \ref{observed parameter}. Comparing the observed properties at different bands, we find that the center of the cores has little difference but the deconvolved core size at Band 6 is always larger than that at Band 7. A possible reason is that the maximum recoverable scale at Band 6 is larger than that at Band 7, which can recover more extended structures.

\begin{table*}
	\centering
	\caption{\centering{Core Observed Properties}}
	\label{observed parameter}
	\begin{tabular}{ccccccccccccc} 
		\hline
		\hline

Core &  RA(J2000) &  Dec(J2000) & $\Theta_{maj}^{a}$ & $\Theta_{min}^{b}$ & PA & $F_\mathrm{peak}$ & $F_\mathrm{core}$ & $R_{\mathrm{eff}}^{c}$ & $R_{\mathrm{eff(dec)}}^{d}$ & RMS Level\\
 & (h:m:s) & (d:m:s) & ($^{\prime \prime}$) & ($^{\prime \prime}$) & (deg) & (Jy beam$^{-1}$) & (Jy) & (AU) & (AU) & (mJy beam$^{-1}$)\\
\hline
Band 7 & & && & & & \\
\hline
ORS-2\underline{\hbox to 0.2cm{}}P1 & 5:41:44.30	&  -1:55:41.3 &	 9.3 & 4.8 &  92 &  2.0 & 4.2 & 2800 & 2000 & 27 \\
ORS-3\underline{\hbox to 0.2cm{}}P1 & 5:41:42.60	&  -1:54:07.9 &	 13.7 & 8.0 &  163 &  0.4 & 2.1 & 4300 & 3300 &9 \\
ORS-3\underline{\hbox to 0.2cm{}}P2 & 5:41:41.54	&  -1:53:49.3 &	 9.8 & 6.1 &  105 &  0.2 & 0.6 & 3200 & 2400 & 9 \\
ORS-4\underline{\hbox to 0.2cm{}}P1 & 5:35:15.78	&  -5:20:39.7 &	 8.5 & 6.2 &  127 &  0.5 & 1.1 & 3000 & 1800 & 11 \\
ORS-4\underline{\hbox to 0.2cm{}}P2 & 5:35:16.74	&  -5:20:43.6 &	 10.4 & 7.0 &  55 &  0.3 & 1.0 &  3500 & 2800 & 11 \\
ORS-5\underline{\hbox to 0.2cm{}}P1 & 5:35:25.30	&  -5:24:32.0 &	 14.2 & 8.7 &  58 &  0.2 & 0.9 & 4600 & 4100 & 4 \\
ORS-7\underline{\hbox to 0.2cm{}}P1 & 5:35:22.18	&  -5:25:04.5 &	 10.9 & 7.2 &  82 &  0.1 & 0.2 & 3700 & 3100 & 4 \\
ORS-8\underline{\hbox to 0.2cm{}}P1 & 5:35:17.98	&  -5:21:29.1 &	 33.0 & 6.1 &  52 &  0.2 & 2.3 & 5800 & 4700 & 8 \\
ORS-9\underline{\hbox to 0.2cm{}}P1 & 5:35:14.52	&  -5:18:46.5 &	 16.9 & 8.6 &  41 &  0.2 & 1.5 & 5000 & 4300 & 6 \\
\hline
Band 6 & & && & & & \\
\hline
ORS-2\underline{\hbox to 0.2cm{}}P1 & 5:41:44.30	&  -1:55:41.2 &	 11.6 & 6.1 &  91 &  1.4 & 2.8 & 3500 & 2500 & 11 \\
ORS-3\underline{\hbox to 0.2cm{}}P1 & 5:41:42.60	&  -1:54:07.9 &	 15.7 & 9.8 &  157 &  0.2 & 1.1 & 5100 & 3900 & 8 \\
ORS-3\underline{\hbox to 0.2cm{}}P2 & 5:41:41.61	&  -1:53:49.3 &	 12.7 & 6.8 &  112 &  0.2 & 0.4 & 3800 & 2800 & 8 \\
ORS-4\underline{\hbox to 0.2cm{}}P1 & 5:35:15.82	&  -5:20:40.2 &	 13.5 & 7.9 &  120 &  0.3 & 0.8 & 4300 & 3200 & 15 \\
ORS-4\underline{\hbox to 0.2cm{}}P2 & 5:35:16.64	&  -5:20:43.5 &	 13.1 & 8.4 &  67 &  0.2 & 0.6 &  4300 & 3400 & 15 \\
ORS-5\underline{\hbox to 0.2cm{}}P1 & 5:35:25.31	&  -5:24:32.2 &	 19.9 & 10.9 &  54 &  0.1 & 0.8 & 6100 & 5300 & 5 \\
ORS-7\underline{\hbox to 0.2cm{}}P1 & 5:35:22.30	&  -5:25:06.1 &	 14.8 & 12.3 &  48 &  0.1 & 0.3 & 5600 & 4800 & 4 \\
ORS-8\underline{\hbox to 0.2cm{}}P1 & 5:35:18.29	&  -5:21:25.0 &	 35.0 & 8.5 &  51 &  0.2 & 1.4 & 7100 & 5700 & 5 \\
ORS-9\underline{\hbox to 0.2cm{}}P1 & 5:35:14.46	&  -5:18:48.0 &	 24.9 & 10.9 &  29 &  0.1 & 1.0 & 6800 & 5600 & 4 \\

\hline
	\end{tabular}
	\begin{tablenotes}
        \footnotesize
        \item $^a$ Beam-convolved FWHM of major axis. 
        \item $^b$ Beam-convolved FWHM of minor axis. 
        \item $^c$ Beam-convolved effective core radius. 
        \item $^d$ Effective core radius deconvolved from the synthesized beam.

      \end{tablenotes}
\end{table*}

\subsection{Mass Estimation}
\par The gas masses of the cores can be derived from the integrated flux at Band 7 because of its higher spatial resolution. Assuming a single dust temperature and optically thin dust emission, we can estimate the core mass following the equation
\begin{equation}
    M_{\text {gas }}=\eta \frac{F_{\nu} d^{2}}{B_{\nu}\left(T_{d}\right) \kappa_{\nu}}
\end{equation}
where $\eta=100$ is the  gas-to-dust ratio, $F_{\nu}$ is the integrated flux at Band 7, $d$ is the source distance, $B_{\nu} \, (T_{d})$ is the Planck function at the dust temperature, and $\kappa_{\nu}$ represents the dust opacity. We adopt the value of 1.37 $\mathrm{cm^{2} \ g^{-1}}$ from \citet{OH_94} for dust opacity at 1.0 mm with a volume density of $10^6$ $\mathrm{cm}^{-3}$. \citet{Jiao_2022} obtained high-resolution dust temperature map ($\sim~ 10^{\prime\prime}$) of Orion A using the J-comb algorithm which combines the high and low-resolution images linearly. For the cloud cores located on this dust temperature map, we calculated the average dust temperature within the cloud cores to determine their mass. For cloud cores not covered by this dust temperature map, we applied the average dust temperature from the dust temperature map using the sophisticated SED model fitting at the angular resolution of $36^{\prime \prime}$ \citep{Lombardi_2014}. The derived gas masses are listed in Table \ref{Physical parameter}, ranging from 0.2 $\mathrm{M_{\odot}}$ to 3.5 $\mathrm{M_{\odot}}$.

\par The applied dust temperature are all higher than 30 K. From Figure \ref{fig:cores} we can see that our sources are located in the regions with strong star-forming activities, indicating that the average dust temperature may be affected by the surrounding environments. Considering the possible overestimation in dust temperature, we calculate the core masses at 15 K, a typical dust temperature for IRDCs. In addition, we also calculate the results at 10 K as the upper limits for the core masses, and the maximum derived mass will be 21.4 $\mathrm{M_{\odot}}$. 

\begin{figure*}[bt!]
    \centering
    \includegraphics[width=1.0\linewidth]{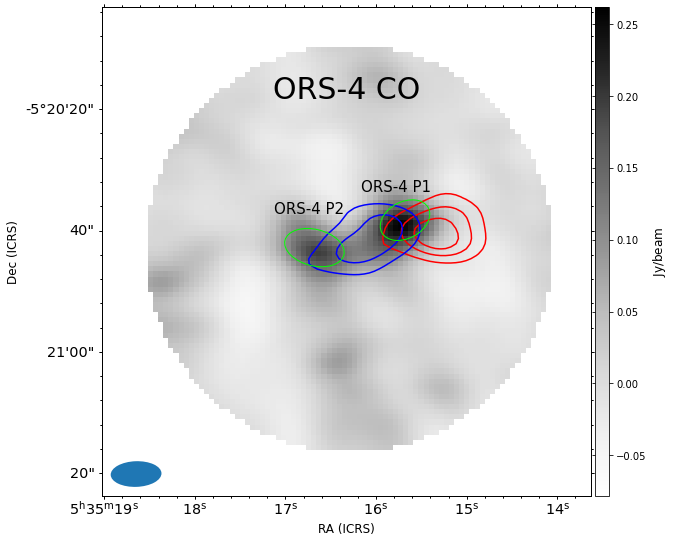}
    \caption{CO outflow of ORS-4\underline{\hbox to 0.2cm{}}P1. The grayscale background image shows the ACA Band 6 continuum emission. The blue ellipse in the bottom-left represents the synthesized beam size of the continuum. The green ellipse marks the position of the cores. The blue and red contours show the blueshifted and redshifted CO emission integrated over [-19, 10] km s$^{-1}$ and [12, 41] km s$^{-1}$, at levels of [40, 80, 120, 160]$\mathrm{~K} \mathrm{~km} \mathrm{~s}^{-1}$. }
    \label{fig:outflow}
\end{figure*}

\begin{table*}
	\centering
	\caption{\centering{Core Physical and Chemical Properties}}
	\label{Physical parameter}
	\begin{tabular}{ccccccccccccc} 
		\hline
		\hline

Core &  $T_d$ &$\rm{Ref\,}^a$ & $M_{\mathrm{gas}}$ & $M_{T_{d}=15 \mathrm{K}}$ &$M_{T_{d}=10 \mathrm{K}}$  &   $N_{\mathrm{N}_{2} \mathrm{D}^{+}}\,^b$  &  $N_{\mathrm{N}_{2} \mathrm{H}^{+}}\,^b$   &   $D_{\mathrm{frac}}$ & Outflow & $\rm{H_2CO}^c$ & $T_{\rm{kin}}$\\
 & (K) & & ($\mathrm{M_{\odot}}$) & ($\mathrm{M_{\odot}}$) & ($\mathrm{M_{\odot}}$)   &  $(10^{10} \ \mathrm{cm}^{-2})$  &  $(10^{11} \ \mathrm{cm}^{-2})$  &   (\%)  &   & (SNR) & (K)\\
\hline
ORS-2\underline{\hbox to 0.2cm{}}P1 & 35 & 1 & 3.5 & 10.9& 21.4  & $<$5.9  & 177 (0.4) & $<$0.3  &  \ding{55} &  28 & 40\\
ORS-3\underline{\hbox to 0.2cm{}}P1 & 33 & 1 & 1.9& 5.6 & 11.0 &  $<$4.7   &  
 203 (0.4)  &  $<$0.2  & \ding{55}  &  44 &  62\\
ORS-3\underline{\hbox to 0.2cm{}}P2 & 39 & 1 & 0.5 & 1.7 & 3.3 &  $<$5.8   &  
 145 (0.5)   &  $<$0.4 & \ding{55}  &  24  &  73\\
ORS-4\underline{\hbox to 0.2cm{}}P1 & 32 & 2 & 1.1 & 3.0 & 5.8&   $<$5.4  &  
 161 (0.5)   &  $<$0.3  &  $\checkmark$ &  22 & 66\\
ORS-4\underline{\hbox to 0.2cm{}}P2 & 32 & 2 & 0.9 & 2.6 & 5.2&  $<$5.2   &
 123 (0.3)   & $<$0.5 &  \ding{55}  &  36 & 139\\
ORS-5\underline{\hbox to 0.2cm{}}P1 & 39 & 2 & 0.7 & 2.3 & 4.6 &  $<$3.3  &  
 $<$3.1   & --- & \ding{55}  &  34 &  62\\
ORS-7\underline{\hbox to 0.2cm{}}P1 & 41 & 2 & 0.2 & 0.6 & 1.2 &   $<$4.5  &  
 $<$3.1   & ---  & \ding{55}  &  14  &  163\\
ORS-8\underline{\hbox to 0.2cm{}}P1 & 37 & 2 & 1.8 &5.9 & 11.7 &  $<$3.1   &  
 131 (0.4)   & $<$0.2  & \ding{55}  &  35 & 34 \\
ORS-9\underline{\hbox to 0.2cm{}}P1 & 31 & 2 & 1.5 & 4.0& 7.9 &  13.2 (1.1)   &  
 82 (0.3)   & 1.6  & \ding{55}  &  31 & 42\\

\hline
	\end{tabular}
	\begin{tablenotes}
        \footnotesize
        \item $^{a}$ Reference: (1) \citet{Lombardi_2014}. (2) \citet{Jiao_2022}.
        \item $^{b}$ The upper limit is set to be 10$\sigma$ for $N_{\mathrm{N}_{2} \mathrm{H}^{+}}$, and 3$\sigma$ for $N_{\mathrm{N}_{2} \mathrm{D}^{+}}$.
        \item $^{c}$ $\rm{H_2CO}$ transition listed here is $3_{2,2}-2_{2,1}$.

      \end{tablenotes}
	
\end{table*}

\subsection{Evolutionary Stages}
\subsubsection{Outflow Signatures}

\par Ordered, bipolar outflow activities are widely used to identify the star formation activities (e.g., \citealt{Beuther_2002,Zhang_2005}). In this study, we employed CO (2-1) and SiO (5-4) to identify outflow signatures in our cores. Although the CO emission shows complex patterns, the bipolar outflows could be identified as pairs of linear emission structures. We examined the blueshifted and redshifted components of CO and SiO emission surrounding the cores and visually inspected the corresponding channel maps. The identified outflow overlaid on continuum map is shown in Figure \ref{fig:outflow}. We only identify one bipolar outflow structure of CO emission in ORS-4\underline{\hbox to 0.2cm{}}P1, indicating that most sources do not exhibit clear evidence of star formation activities.


\subsubsection{Deuterium Fraction}

\begin{figure}[tb!]
    \centering
    \includegraphics[width=0.48\textwidth]{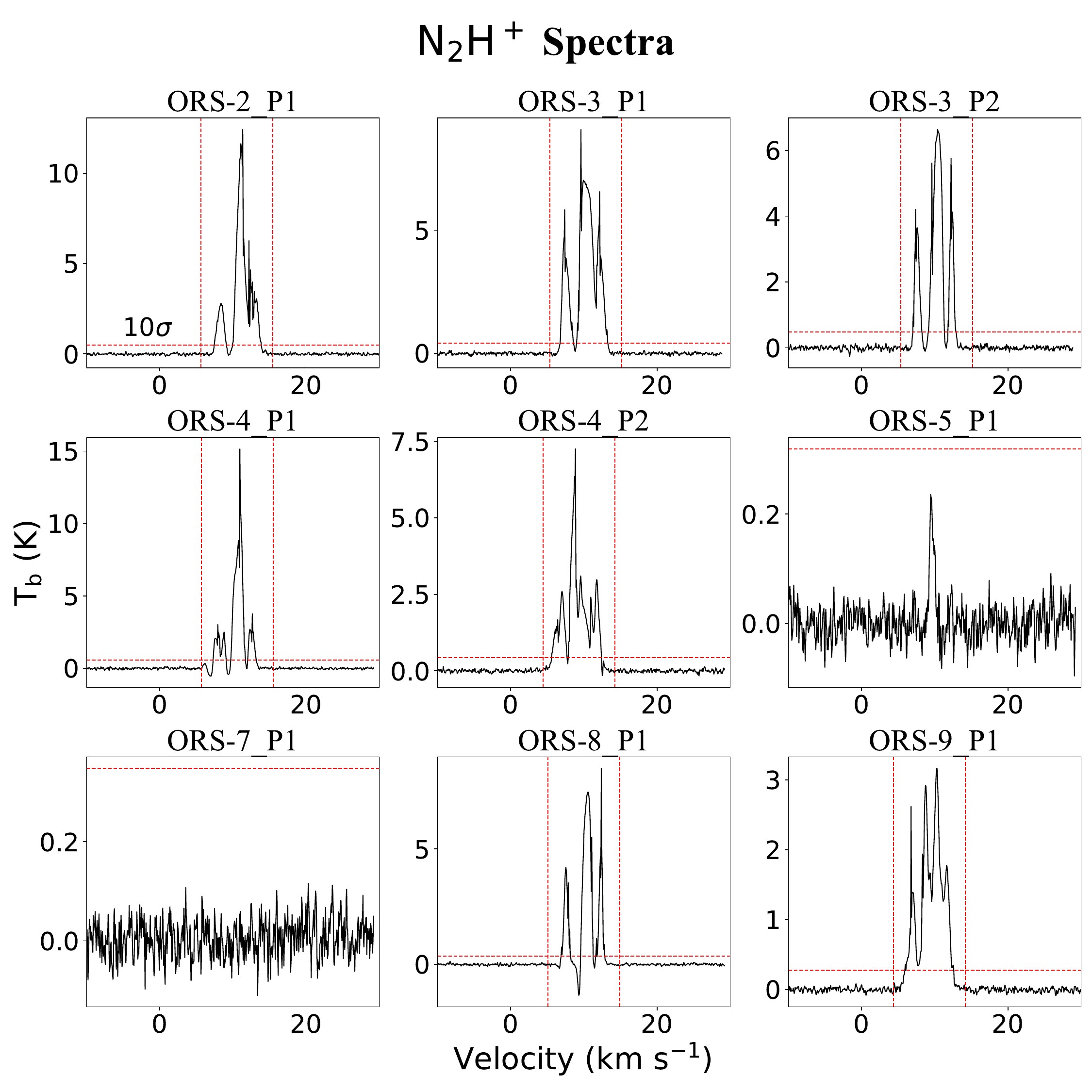}
    \hspace{0in}
    \includegraphics[width=0.48\textwidth]{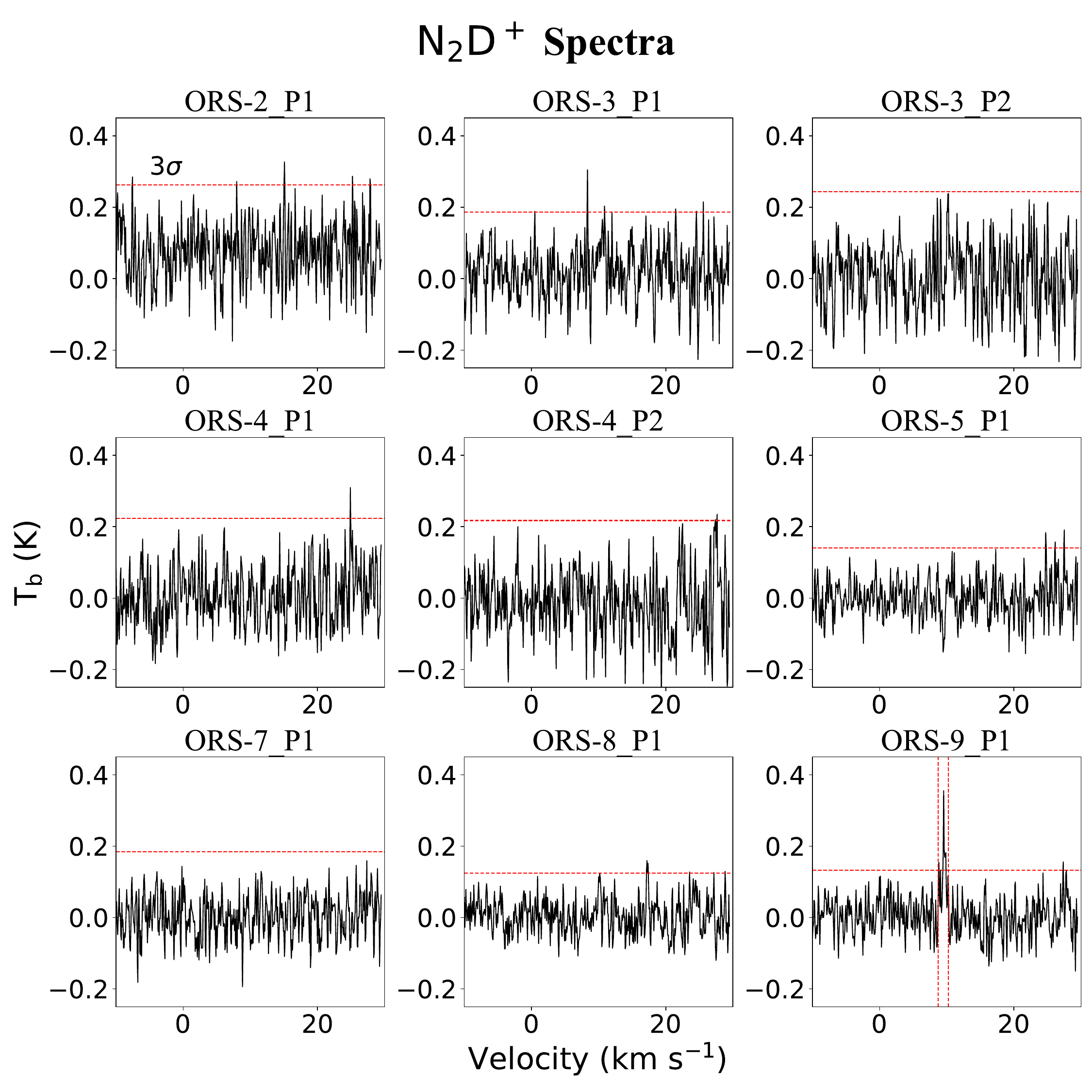}
    \caption{$\mathrm{N_2H^+}$ and $\mathrm{N_2D^+}$ spectra in nine cores. Left: $\mathrm{N_2H^+}$ spectra. The red horizontal dotted lines represent 10$\sigma$ limits, and the red vertical lines represent the integrating range. Right: $\mathrm{N_2D^+}$ spectra. The red horizontal dotted lines represent 3$\sigma$ limits, and the red vertical lines represent the integrating range.}
    \label{fig:deuteration_spectra}
\end{figure}

\begin{figure*}[bt!]
    \centering
    \includegraphics[width=1.0\linewidth]{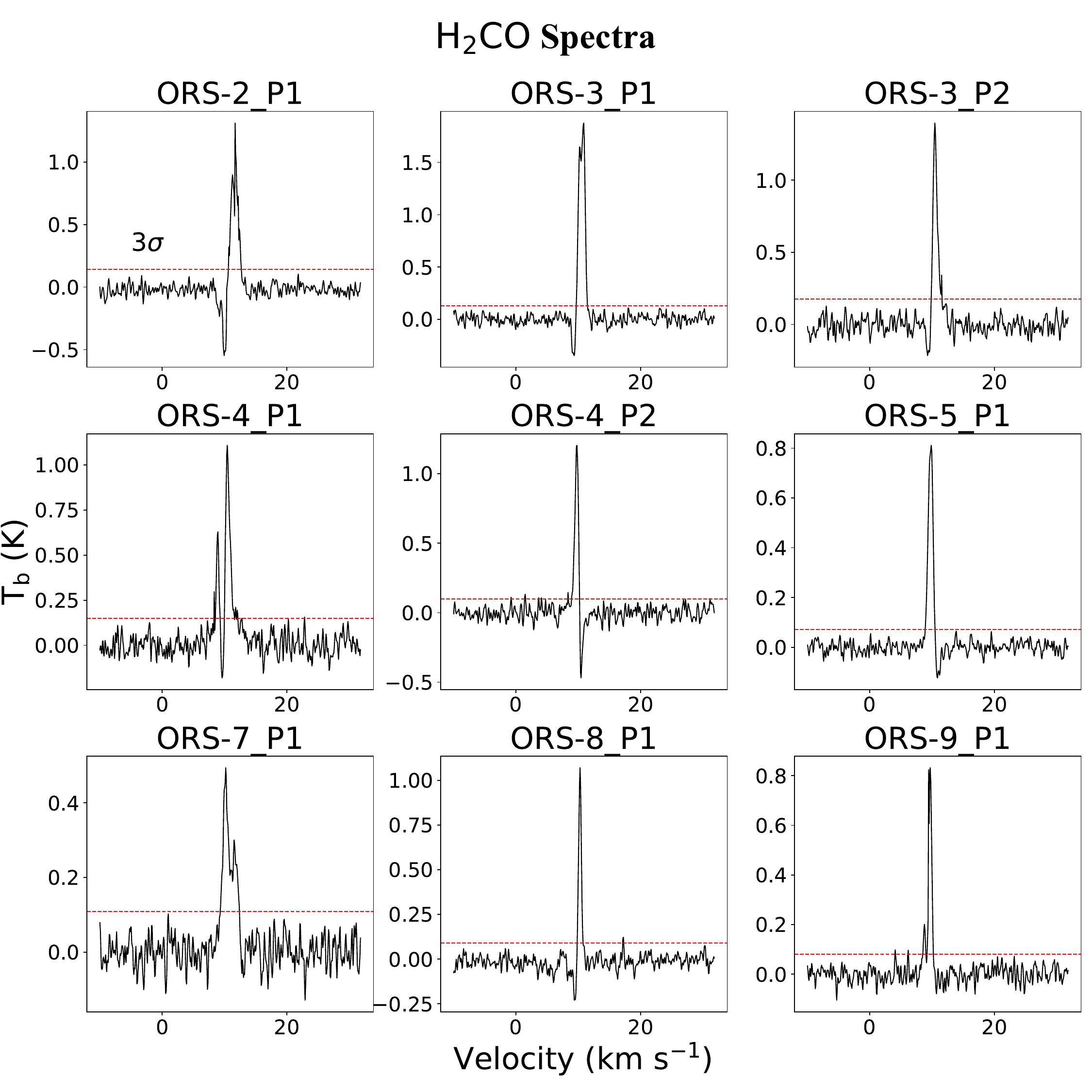}
    \caption{$\rm{H_2CO~(3_{2,2}-2_{2,1})}$ spectra in nine cores. The red dotted lines represent 3$\sigma$ limits.}
    \label{fig:h2co}
\end{figure*}

Due to the feedback of star formation activities, the chemical abundance will undergo temporal evolution. Previous studies established a correlation between the evolution of prestellar cores and the deuterium fraction (especially for $\mathrm{N}_{2} \mathrm{D}^{+}$/$\mathrm{N}_{2} \mathrm{H}^{+}$), which predicts that the deuterium fraction is highest in the prestellar stage and then decreases after the protostellar stage (\citealt{Caselli_2002,Fontani_2011}). The spectra of $\mathrm{N_2H^+}$ and $\mathrm{N_2D^+}$ within the FWHM of nine cores are shown in Figure \ref{fig:deuteration_spectra}. The evolutionary stage of the nine cores identified in this work was analyzed by calculating the deuterium fraction of $\mathrm{N_2H^+}$ using two-band data. Assuming that the excitation temperature equals to 15 K, the beam filling factor is 1, optically thin emission, negligible background temperature, and applying the Rayleigh-Jeans approximation and local thermodynamic equilibrium (LTE) conditions, we can calculate the total column density of $\mathrm{N}_{2} \mathrm{D}^{+}$ and $\mathrm{N}_{2} \mathrm{H}^{+}$ within core size using the following equation:

\begin{equation}
     N_{\mathrm{tot}}={(\frac{3 k_B}{8 \pi^{3} \nu S \mu_d^{2} }) \  (\frac{Q_{\mathrm{rot}}}{g_{J} g_{K} g_{I}}) \exp({ \frac{E_{u}}{k_B T_{e x}}}) \int {T_\mathrm{B} d v}}
\end{equation}
where ${k_B}$ is the Boltzmann constant, $\nu$ is the rest frequency of the line, $S$ is the line strength, $\mu_d$ is the permanent dipole moment of the molecule, $Q_\mathrm{rot}$ is the partition function of the molecule, $g_i$ are the degeneracies, $E_u$ is the energy of the upper energy level, $T_{ex}$ is the excitation temperature. To ensure the reliability of the results, spectra must exhibit at least five continuous velocity channels with a brightness temperature higher than 10 $\sigma$ for $\mathrm{N_2H^+}$ and higher than 3 $\sigma$ for $\mathrm{N_2D^+}$. We only detect $\mathrm{N_2D^+}$ in one core but detected $\mathrm{N_2H^+}$ in seven cores. The upper limits are calculated using the equation: $N_{upper} = \rm{SNR} \times \sigma_{N}$, where SNR is 10 for both $\mathrm{N_2H^+}$ and $\mathrm{N_2D^+}$. Here, $\sigma_{N}$ is the uncertainty of the column density, and the integrated line emission is given by $\sqrt{\rm{N}}\sigma_{c}$, where $\sigma_{c}$ is the rms noise level of individual channels and N is the number of integrated channels. We used the RADEX online tool to estimate the optical depth of $\mathrm{N_2H^+}$. The derived optical depth for each hyperfine component ranges from $10^{-3}$ to $5 \times 10^{-1}$, indicating that the optically thin assumption is reasonable. Deuterium fraction is calculated by $D_{\rm{frac}}=N_{N_2D^+}/N_{N_2H^+}$, and all  derived values of $D_{\rm{frac}}$ is lower than 2\%. The  results are listed in Table \ref{Physical parameter}. Such deuterium fraction is low compared to that in high-mass star-forming regions at early evolutionary stages (usually $>$15\%; \citealt{Kong_2016,Jiao_2023}). Additionally, previous studies reported that the detection rate and line intensities of $\rm{H_2CO}$ are low in prestellar cores but high in protostellar cores \citep{Li_2022}. The $\rm{H_2CO~(3_{2,2}-2_{2,1})}$ line, with its high upper energy, is a good tracer for warm cores \citep{Tang_2021,Gieser_2021}. The core-averaged spectra of $\rm{H_2CO~(3_{2,2}-2_{2,1})}$ line are shown in Figure \ref{fig:h2co}. As shown in Table \ref{Physical parameter}, we detect $\rm{H_2CO~(3_{2,2}-2_{2,1})}$ with an SNR higher than 10 in all cores. We estimate the kinetic temperature assuming Local Thermodynamic Equilibrium (LTE) and optically thin lines using the following equation \citep{Mangum_1993}:
\begin{equation}
    T_{\mathrm{kin}}=\frac{47.1}{\ln (0.556 \frac{I(3_{03}-2_{02})}{I(3_{22} -2_{21})})}
\end{equation}
The derived kinetic temperature ranges from 30 to 140 K, as shown in Table \ref{Physical parameter}, suggesting that the dust temperatures of the cores are not very low. 

Although molecular outflows are rarely detected, the low deuterium fraction of $\mathrm{N_2H^+}$ and high kinetic temperature indicate that the cores in our sample are more likely to be at protostellar stages rather than prestellar stages. The low detection rate of outflows in our sample could be due to two reasons: previous studies have found that some protostars don’t have detectable outflows (e.g., 38\% in the ASHES survey; \citealt{Li_2022}), and our ACA observations might miss some potential outflows due to limited resolution, sensitivity, and image size (see Figure \ref{fig:outflow_other_cores} for more details). This supports using 15 K as a reliable lower limit for the dust temperature. Consequently, the derived core mass will not exceed 10.9 $\mathrm{M_{\odot}}$.

\section{Discussion}
\label{sec:discussions}

\begin{figure*}[bt!]
    \centering
    \includegraphics[width=1.0\linewidth]{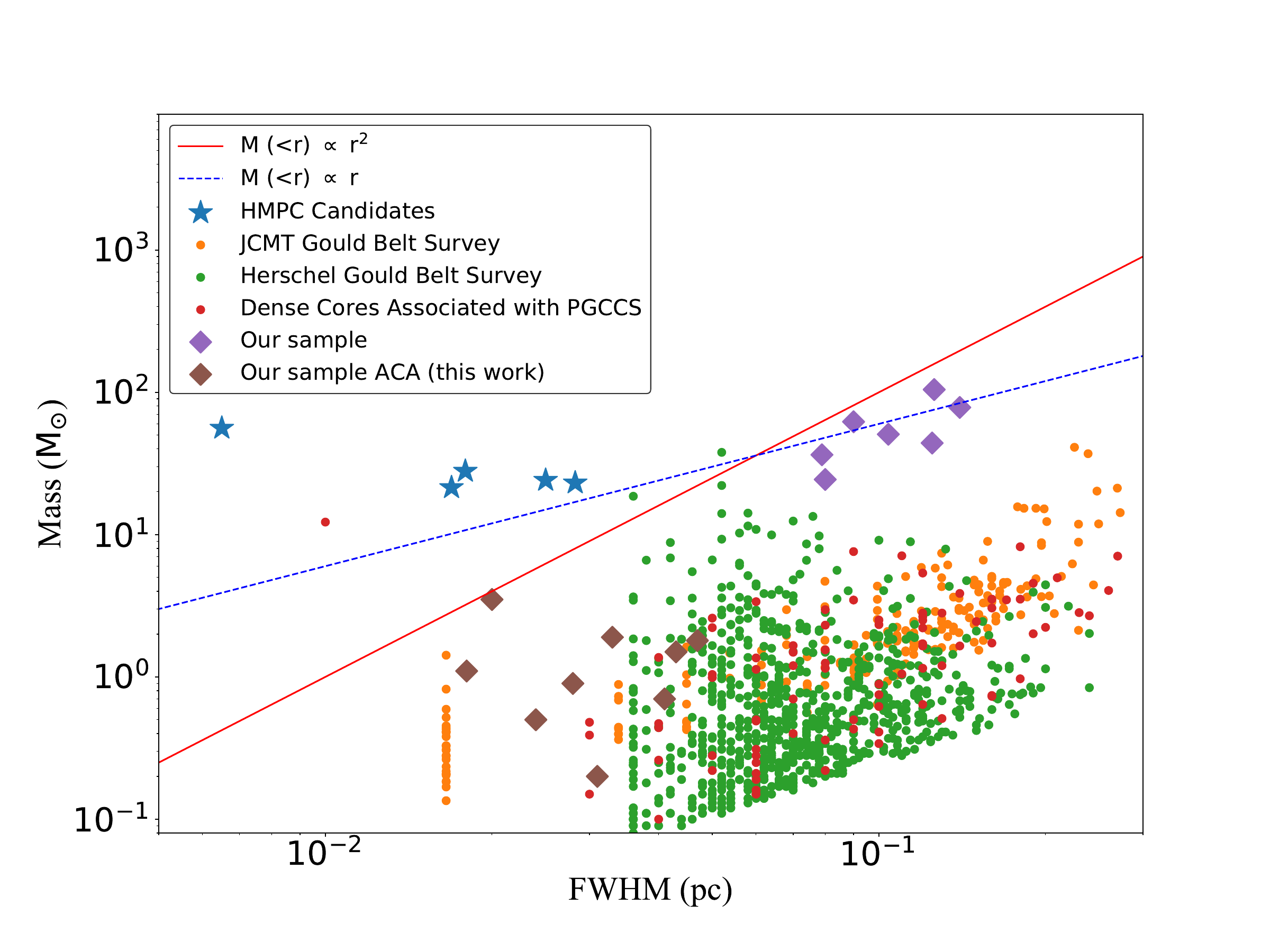}
    \caption{Mass-size diagram of the prestellar cores in Orion. The blue stars represent the known high-mass prestellar core candidates in the galactic plane for comparison \citep{Bontemps_2010,Wang_2014,Nony_2018,Nony_2023,Barnes_2023}. The purple and brown diamonds represent the core properties in this paper observed by JCMT \citep{Nutter_2007} and ACA. The orange, green, and red dots depict the prestellar/starless cores identified by \citet{Salji_2015}, \citet{Konyves_2020}, and \citet{Yi_2018}, respectively. The lines illustrate the mass-size scalings of the clumps, following a turbulence dominated (M  $\propto \ \rm{r^2}$) or gravity dominated (M $\propto$ r) density structure \citep{Motte_2018}. }
    \label{fig:sample}
\end{figure*}

Based on our results, we did not find high-mass prestellar cores in our sample. In this section, we will discuss whether such cores exist in the Orion region. Recent single-dish surveys have systematically explored the physical properties of dense cores within the Orion Nebula. \cite{Salji_2015} used JCMT observations of the Orion A North molecular cloud and identified 210 prestellar cores. \cite{Yi_2018} compared the properties of Planck Galactic Cold Clumps (PGCCs) in the $\lambda$ Orionis cloud with those in the Orion A and B clouds using JCMT data. \cite{Kirk_2016} provided an initial examination of dense cores in Orion B from the JCMT Gould Belt Legacy Survey, while \cite{Konyves_2020} conducted a comprehensive study of dense cores in the Orion B molecular cloud complex using data from the Herschel Gould Belt Survey (HGBS), identifying hundreds of prestellar cores. We compile the results of these studies alongside our own in Figure \ref{fig:sample}. The red solid line and the blue dotted line represent turbulence-dominated and gravity-dominated mass concentration mechanisms, respectively. Only one prestellar core in HGBS is above the blue dotted lines, and this core is actually the ORS-3 in our sample, fragmenting to two small cores at 0.01 pc resolution. The mass and FWHM size of ORS-3 reported in the JCMT survey \citep{Nutter_2007},  the Herschel Gould Belt survey
 \citep{Konyves_2020}, and this work are 78.3 $M_{\odot}$ and 0.14 pc, 37.85 $M_{\odot}$ and 0.05 pc, and 1.9/0.5 $M_{\odot}$ and 0.03/0.02 pc, respectively. Different observations probe different scales. Nevertheless, 
 regardless of the underlying mass concentration mechanism, the lack of high-mass prestellar cores is common in these researches.

\par In addition, higher resolution observations were performed toward sub-regions in Orion, revealing the initial conditions of star formation at smaller scales (e.g., \citealt{Sahu_2021,Sato_2023}). On the one hand, many prestellar cores found in single-dish observations are identified as being at the protostellar stage with higher resolution observations \citep{Dutta_2020}. On the other hand, the presence of smaller-scale fragmentation \citep{Palau_2015, Fielder_2024} and the extended emission surrounding the core \citep{Palau_2018} can result in core masses that are significantly smaller than those derived from single-dish observations. It further proves the scarcity of high-mass prestellar cores in Orion. This deduction is to put the new findings into context of the field. Orion GMC, as a well studied massive star-forming regions, may not have massive starless cores. It poses a significant challenge to the existence of high-mass prestellar cores.

\par Outside of Orion, numerous studies have attempted to search the high-mass prestellar cores by targeting infrared dark clouds (IRDCs) in the galactic plane (e.g., \citealt{Zhang_2009,qz15,Wang_2014,Kong_2017, Sanhueza_2019, Svoboda_2019}), but only few promising HMPC candidates have been found \citep{Wang_2014, Barnes_2023}. Yet, \citet{Motte_2018_NA} reported a top-heavy core mass function in evolved protostellar clusters, indicating significant gas accumulation in proclusters. 
\cite{KongS2021} found evidence of core growth in IRDC G28.34+0.06, by censusing its core population and differentiating pre- and protostellar cores.
Both the ALMA-IMF \citep{Nony_2023} and ASSEMBLE \citep{Xu_2024} projects presented prestellar cores with higher masses than those found in IRDCs, suggesting the possible existence of high-mass prestellar cores in protoclusters. Potential HMPC candidates, such as W43-MM1$\#$6 \citep{Nony_2018} and W43-MM1$\#$134 \citep{Nony_2023}, have been reported in protoclusters. Additionally, another HMPC candidate, CygXN53-MM2 \citep{Bontemps_2010}, was identified in CygnusX, the second nearest high-mass star-forming region. We have illustrated these known HMPC candidates in Figure \ref{fig:sample} for comparison. The scarcity of HMPC candidates either challenges the existence of high-mass prestellar cores or indicates a very short lifetime for this phase. Moreover, in recent years, several promising high-mass prestellar core candidates were excluded in follow-up investigations where fragmentation or outflow activities were identified (e.g., \citealt{Tan_2016, Cyganowski_2022, Mai_2024}). Further detailed studies are necessary to determine whether the remaining candidates are indeed massive, starless cores.


\section{Conclusions} 
\label{sec:summary}

In this work, we conducted ALMA ACA Band 6 and Band 7 continuum and line observations targeting seven selected massive starless core candidates in the Orion region. We identified nine dense cores across both bands, with masses ranging from 0.2 $\rm{M_{\odot}}$ to 3.5 $\rm{M_{\odot}}$. Using various dust temperatures, the maximum core mass was found to be less than 11 $\rm{M_{\odot}}$. No high-mass prestellar cores were found in our sample, and this absence is consistent with previous surveys in Orion. These findings challenges the existence of high-mass prestellar cores. 

To further investigate this, detailed studies of the remaining high-mass prestellar core candidates are essential, providing deeper insights into the initial conditions and processes involved in high-mass star formation.


\begin{acknowledgments}
We are grateful to an anonymous referee for the constructive comments that helped us improve this
paper. This work has been supported by the National Science Foundation of China (12041305, 12033005),
the National Key Research and Development Program of China (2022YFA1603100),
the Tianchi Talent Program of Xinjiang Uygur Autonomous Region, and
the China-Chile Joint Research Fund (CCJRF No.\,2211). CCJRF is provided by Chinese Academy of Sciences South America Center for Astronomy (CASSACA) and established by National Astronomical Observatories, Chinese Academy of Sciences (NAOC) and Chilean Astronomy Society (SOCHIAS) to support China-Chile collaborations in astronomy. F.W.X acknowledges the funding from the European Union's Horizon 2020 research and innovation programme under grant agreement No 101004719 (ORP).
This paper makes use of the following ALMA data: ADS/JAO.ALMA\#2019.2.00094.S ALMA is a partnership of ESO (representing its member states), NSF (USA) and NINS (Japan), together with NRC (Canada), MOST and ASIAA (Taiwan), and KASI (Republic of Korea), in cooperation with the Republic of Chile. The Joint ALMA Observatory is operated by ESO, AUI/NRAO and NAOJ. 
\end{acknowledgments}

%

\vspace{5mm}


\software{CASA \citep{McMullin_2007}; astropy \citep{Astropy_2013}
          }



\appendix

\section{CO emission}

\begin{figure*}[bt!]
    \centering
    \includegraphics[width=1.0\linewidth]{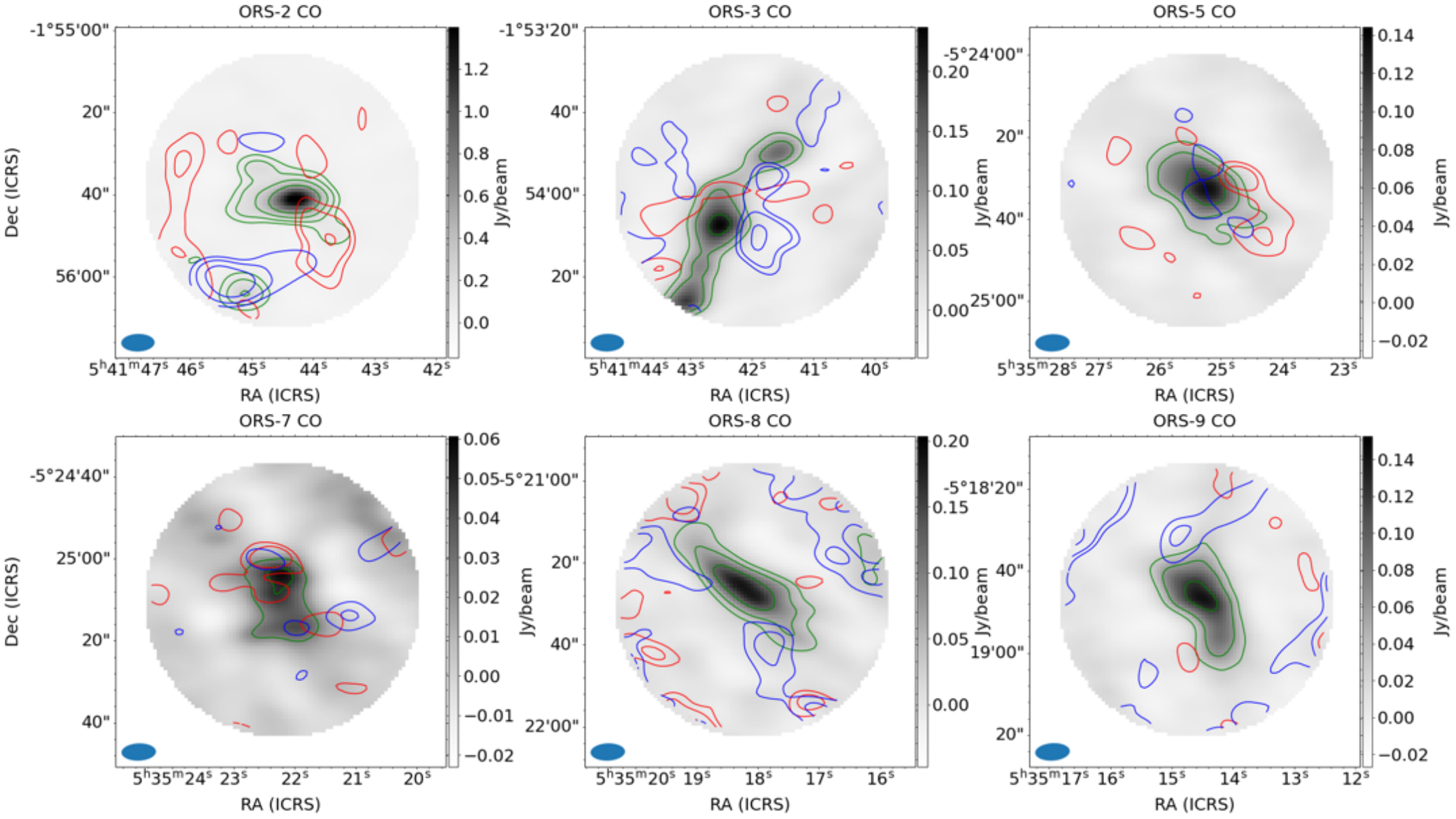}
\caption{CO outflow plots for all cores except ORS-4. The grayscale background image shows the ACA Band 6 continuum emission without primary beam correction. The blue ellipse in the bottom left represents the synthesized beam size of the continuum. Green contours indicate the continuum emission at levels of [6, 12, 24, 48]$\sigma$. The blue and red contours represent the blueshifted and redshifted CO emission without primary beam correction, respectively. Contour levels vary due to different emission levels.}
    \label{fig:outflow_other_cores}
\end{figure*}

Figure \ref{fig:outflow_other_cores} shows the integrated blueshifted and redshifted CO emission of all cores except ORS-4. The CO emissions appear to be messy, and no clear outflows can be identified. However, the possibility of potential outflows cannot be ruled out.

\bibliography{sample631}{}
\bibliographystyle{aasjournal}



\end{document}